\documentclass[aps,prl,twocolumn,showpacs,superscriptaddress,groupedaddress]{revtex4} 

\usepackage{graphicx}
\usepackage{dcolumn} 
\usepackage{bm}
\usepackage{amsmath}
\usepackage{amssymb}
\usepackage{braket}

\newcommand{\up}{\uparrow}
\newcommand{\down}{\downarrow}

\newcommand{\ep}{\varepsilon}
\def\lesssim{\ \raise.3ex\hbox{$<$}\kern-0.8em\lower.7ex\hbox{$\sim$}\ }
\def\gesim{\ \raise.3ex\hbox{$>$}\kern-0.8em\lower.7ex\hbox{$\sim$}\ }

\begin{document}
\title{Proposed Fermi-surface reservoir-engineering and application to realizing unconventional Fermi superfluids}
\author{Taira Kawamura}
\affiliation{Department of Physics, Keio University, 3-14-1 Hiyoshi, Kohoku-ku, Yokohama 223-8522, Japan}
\author{Ryo Hanai}
\affiliation{Asia Pacific Center for Theoretical Physics, Pohang 37673, Korea}
\affiliation{Department of Physics, POSTECH, Pohang 37673, Korea}
\author{Yoji Ohashi}
\affiliation{Department of Physics, Keio University, 3-14-1 Hiyoshi, Kohoku-ku, Yokohama 223-8522, Japan}
\date{\today}
\begin{abstract}
We theoretically propose an idea based on reservoir engineering to process the structure of a Fermi edge to split into {\it multiple} Fermi edges, so as to be suitable for the state which we want to realize. When one appropriately tunes the chemical-potential difference between two reservoirs being coupled with the system, the system is shown to be in the non-equilibrium steady state with the momentum distribution having a two-edge structure. We argue that these edges play similar roles to two Fermi surfaces, which can be designed to realize exotic quantum many-body states. To demonstrate this, we consider a model driven-dissipative two-component Fermi gas with an attractive interaction as a paradigmatic example and show that it exhibits an unconventional Fermi superfluid. While the superfluid order parameter of this state has the same form as that in the Fulde-Ferrell state discussed in metallic superconductivity under an external magnetic field, the former non-equilibrium pairing state is {\it not} accompanied by any spin imbalance. Our proposed reservoir engineering to process the Fermi momentum distribution would provide further possibilities of many-body quantum phenomena beyond the thermal equilibrium case.
\end{abstract}
\maketitle
\par
\textit{Introduction}--\ Fermi surface (FS) is sometimes called ``the face of metal," because of its crucial contribution to system properties. The sharp Fermi surface edge plays central roles in various effects \cite{Kondobook} associated with orthogonal catastrophe \cite{Anderson}, such as the Kondo effect, quantum diffusion, as well as edge-singularity of soft $X$-ray absorption. In Fermi superfluids, FS assists the pair formation (Cooper instability) \cite{Cooper1956}. A nested FS stabilizes the spin- and charge-density-wave states \cite{Grunerbook}.
\par
\begin{figure}[t]
\centering
\includegraphics[width=80mm]{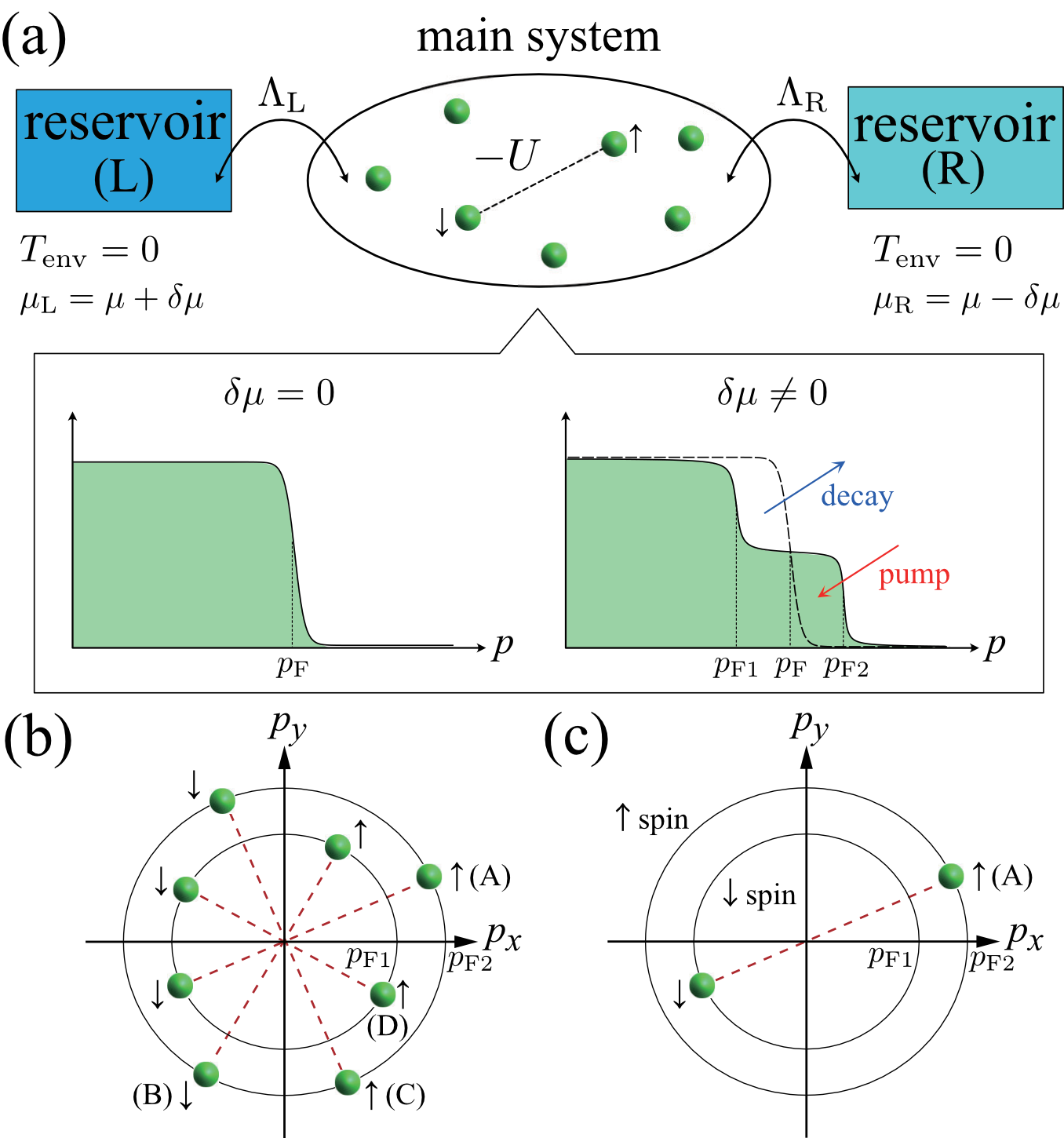}
\caption{(a) Model non-equilibrium driven-dissipative two-component Fermi gas ($\sigma=\up,\down$) with an $s$-wave pairing interaction $-U (<0)$. The non-equilibrium main system is coupled with two reservoirs ($\alpha={\rm L,R}$) consisting of free fermions in the thermal equilibrium state. These reservoirs have the common environment temperature $T_{\rm env}=0$, but different chemical potentials $\mu_{\rm L}=\mu+\delta\mu$ and $\mu_{\rm R}=\mu-\delta\mu$. When the system is in the steady state, the non-zero chemical potential difference ($\delta\mu\neq 0$) produces two edges at $p_{{\rm F}1}$ and $p_{{\rm F}2}$ in the Fermi momentum distribution $n_{{\bm p},\sigma}$. (b) Expected types (A)-(D) of Cooper pairs, when the two edges work like two Fermi surfaces. (c) Pairing structure in the ordinary (thermal equilibrium) FF state under an external magnetic field.}
\label{fig1} 
\end{figure}
\par
Because of these, one can alter properties of a many-fermion system by tweaking the size/shape/topology of FS (which may even trigger a phase transition). Indeed, the above-mentioned Fermi surface effects can be tuned by the temperature, because it smears the sharp Fermi surface edge. An external magnetic field splits $\uparrow$-spin and $\downarrow$-spin Fermi surfaces, which may replace the conventional $s$-wave BCS superconducting state by the Fulde-Ferrell (FF) one \cite{Fulde1964,Takada1969}. Further examples are using pressure \cite{Chu1970, Prieto2006, Glazyrin2013, Xiang2015, Kang2015, Gonnelli2016}, strain \cite{Testardi1970, Martins1978, Hsu2016, Burganov2016, Sunko2019}, and doping \cite{Armitage2002, Kusunose2003, Kaminski2006, Liu2010, Shi2017}. 
\par
Besides the thermal equilibrium case, such ``FS engineering" has also been discussed in the non-equilibrium state \cite{Beaulieu2020, Kirby2020}, where a Lifshitz transition has been observed in the transient dynamics of a highly excited Weyl semimetal. However, the current processing technology is still within simple level, that is, smearing/deforming of FS, and utmost creating FSs through the Lifshitz transition. Even for an external magnetic field, the Fermi surface of each spin component only changes its size.
\par
In this letter, we propose an alternative approach to the FS engineering, which can imprint a multi-step structure on the Fermi momentum distribution $n_{{\bm p},\sigma=\up,\down}$, to explore exotic quantum many-body states that have not been realized/discussed/known in condensed matter physics. 
Our idea assumes a coupled Fermi gas with reservoirs having different values of chemical potentials. In the simplest model shown in Fig. \ref{fig1}(a), for example, the two reservoirs supply fermion to the system up to their respective chemical potentials, giving rise to a two-step structure in $n_{{\bm p},\sigma=\up,\down}$ as its non-equilibrium steady state (NESS). Such a two-step structure of $n_{{\bm p},\sigma=\up,\down}$ has been realized in voltage-biased mesoscopic wires \cite{Pothier1997, Anthore2003}, as well as in carbon nanotubes \cite{Chen2009}. In cold atom physics, the recent experimental development of non-equilibrium techniques \cite{Brantut2012, Krinner2015, Husmann2015, Krinner2017} makes us expect the implementation of ``FS reservoir-engineering" in the near future.
\par
If each edge imprinted on $n_{{\bf p},\sigma}$ works like a Fermi surface (which we show below that it indeed does), it means that one can produce plural Fermi surfaces from one Fermi sphere. Then, in the model case in Fig. \ref{fig1}(a), an $s$-wave attractive interaction $-U$ is expected to produce four types of Cooper pairs (A)-(D) shown in Fig. \ref{fig1}(b), two of which are unconventional.  While (C) and (D) are essentially the same as the ordinary BCS pairing, (A) and (B) are non-zero center-of-mass momentum pairing that is rather close to the unconventional FF state (see Fig. \ref{fig1}(c)) discussed in superconductivity under an external magnetic field \cite{Fulde1964, Takada1969, Matsuda2007}, spin-polarized Fermi gases \cite{Liao2010, hu2006, Chevy2010, Kinnunen2018}, as well as color superconductivity in quantum chromodynamics \cite{Casalbuoni2004}. We recall that the FF state is usually realized in the spin-imbalanced case, where FF Cooper pairs are formed between $\up$-spin fermions around the larger Fermi surface in Fig. \ref{fig1}(c) and $\down$-spin ones around the smaller Fermi surface, as symbolically written as $|{\rm (A)}\rangle=|-{\bm p}_{\rm F1},\down\rangle|{\bm p}_{\rm F2},\up\rangle$. In contrast, the model driven-dissipative Fermi gas in Fig. \ref{fig1}(a) is {\it not} accompanied by any spin imbalance, but each spin component has two ``Fermi edges" at $p_{\rm F1}$ and $p_{\rm F2}$. This leads to the pairing $|{\rm (B)}\rangle=|{\bm p}_{\rm F1},\up\rangle|-{\bm p}_{\rm F2},\down\rangle$, in addition to $|{\rm (A)}\rangle$. In a sense, the non-equilibrium FF-like (NEFF) state may be viewed as a mixture of two FF states under external magnetic fields ${\bm B}$ and $-{\bm B}$.
\par
We note that possible routes to the FF state in the spin-balanced case has been discussed in the literature, where the shift of single-particle energy induced by external current \cite{Doh2006}, a size effect \cite{Vorontsov2009}, an inter-atomic interaction \cite{He2018}, and an artificial field \cite{Zheng2015, Zheng2016, Nocera2017}, have been proposed to realize this unconventional Fermi superfluid. However, these ideas are all in the thermal equilibrium case with the ordinary Fermi distribution function, which is quite different from our idea in the non-equilibrium state.
\par
In what follows, we confirm our scenario by dealing with the model driven-dissipative Fermi gas in Fig. \ref{fig1}(a). We show that our proposed FS reservoir-engineering really produces a two-step structure in $n_{{\bm p},\sigma}$. We then show that this processed momentum distribution causes the NEFF phase transition. The latter result proves that two edges imprinted on $n_{{\bm p},\sigma}$ work like two Fermi surfaces. Throughout this letter, we set $\hbar=k_{\rm B}=1$, and the system volume $V$ is taken to be unity, for simplicity.
\par
\par
\textit{FS reservoir-engineering}--\ The driven-dissipative Fermi gas in Fig. \ref{fig1}(a) is described by the Hamiltonian $H=H_{\rm sys}+H_{\rm env}+H_{\rm t}$, consisting of the main system term $H_{\rm sys}$, the reservoir term $H_{\rm env}$, as well as the tunneling term $H_{\rm t}$ \cite{Kawamura2020JLTP, Kawamura2020}. The reservoir term has the form,
\begin{equation}
H_{\rm env} = \sum_{\alpha={\rm L}, {\rm R}} \sum_{\bm{p}, \sigma} \xi^\alpha_{\bm{p}} c^{\alpha\dagger}_{\bm{p} \sigma} c^{\alpha}_{\bm{p}\sigma},
\end{equation}
where $\alpha={\rm L,R}$ denote the left and right reservoirs, and $c^\alpha_{\bm{p}\sigma}$ is the annihilation operator of a fermion with (pseudo-)spin $\sigma=\up,\down$ in the $\alpha$-reservoir. Each reservoir is assumed to be a free Fermi gas with the kinetic energy $\xi^{\alpha={\rm L,R}}_{\bm{p}}=\ep_{\bm{p}}-\mu_\alpha$, measured from the Fermi chemical potential $\mu_\alpha$, where $\ep_{\bm p}={\bm p}^2/(2m)$ with $m$ being a particle mass. We also assume that the reservoirs are huge compared to the main system and are always in the ground state at $T_{\rm env}=0$. Fermions in these reservoirs thus obey the ordinary Fermi distribution function at $T_{\rm env}=0$, $f(\xi^\alpha_{\bm{p}})=\Theta(-\xi^\alpha_{\bm{p}})$, where $\Theta(x)$ is the step function. 
\par
The tunneling between these reservoirs and the main system is described by $H_{\rm t}=\sum_{\alpha={\rm L,R}}H_{\rm t}^\alpha$, where 
\begin{equation}
H_{\rm t}^\alpha = \sum_{j=1}^{N_{\rm t}} \sum_{\bm{p}, \bm{q}, \sigma}\Big[e^{i[\mu_\alpha t+\bm{p}\cdot \bm{r}^\alpha_j-\bm{q}\cdot \bm{R}_j^\alpha]} \Lambda_\alpha c^{\alpha\dagger}_{\bm{q}\sigma} a_{\bm{p}\sigma}+ {\rm H.c.}\Big].
\label{tunnelH}
\end{equation}
In Eq. (\ref{tunnelH}), particles tunnel between spatial positions $\bm{R}_j^\alpha$ in the $\alpha$-reservoir and $\bm{r}_j$ in the main system ($j=1, \cdots N_{\rm t}\gg 1$). Although the translational invariance of the main system is broken in Eq. (\ref{tunnelH}), this symmetry property recovers by taking spatial averages over ${\bm R}_i^{\alpha}$ and ${\bm r}_i$ \cite{Hanai2016, Hanai2017, Hanai2018}. For simplicity, we set the tunneling matrix elements as $\Lambda_{\rm L}=\Lambda_{\rm R}\equiv \Lambda$. The main system becomes in the non-equilibrium state when $\mu_{\rm L}=\mu+\delta\mu$ and $\mu_{\rm R}=\mu -\delta\mu$ with $\delta\mu \neq 0$.
\par
To grasp effects of $\delta\mu \neq 0$ in more detail, we first consider the simple case when the main system is a two-component {\it free} Fermi gas, that is $H_{\rm sys}=H_{0} = \sum_{\bm{p}, \sigma} \ep_{\bm{p}} a^\dagger_{\bm{p} \sigma} a_{\bm{p} \sigma}$ (where $a_{\bm{p}\sigma}$ is the annihilation operator of a fermion in the main system). In this case, after a long time has passed since the system was connected to the reservoirs, the main system would reach NESS, where the gain and loss of particles in the main system are balanced. The momentum distribution $n_{\bm{p},\sigma}=\braket{a^\dagger_{\bm{p}\sigma} a_{\bm{p}\sigma}}$ in NESS can be evaluated by the ordinary Keldysh Green's function technique \cite{Kawamura2020JLTP, Kawamura2020}, which yields
\begin{equation}
n_{\bm{p},\sigma}=\frac{1}{2}-\frac{1}{2\pi}\sum_{\zeta=\pm}{\rm Tan}^{-1}\left( \frac{\xi_{\bm{p}} +\zeta\delta\mu}{2\gamma} \right). 
\label{eq_fneq}
\end{equation}
Here, $\xi_{\bm{p}}=\ep_{\bm{p}} -\mu$, and $\gamma= \pi N_{\rm t} \rho |\Lambda|^2$ is the quasi-particle damping rate, where $\rho$ is the single-particle density of states in the reservoirs. In obtaining Eq. (\ref{eq_fneq}), we have taken the spatial averages over the tunneling positions, and have ignored the $\alpha(={\rm R}, {\rm L})$ and $\omega$ dependence of $\rho$ \cite{Stefanucci2013}. In the weak-damping limit ($\gamma\to +0$), the momentum distribution in Eq. (\ref{eq_fneq}) has a two-step structure, as $n_{\bm{p},\sigma}=\frac{1}{2}[\Theta(\xi_{\bm{p}}+\delta\mu)+\Theta(\xi_{\bm{p}}-\delta\mu)]$. Thus, it has two edges at $p_{{\rm F}1}=\sqrt{2m(\mu-\delta\mu)}$ and $p_{{\rm F}2}=\sqrt{2m(\mu+\delta\mu)}$ as schematically shown in Fig. \ref{fig1}(a). Although these edges become obscure as $\gamma$ increases, they remain as far as $\gamma \ll \delta\mu$.
\par
\par
\textit{NEFF superfluid instability}--\ We next show that the imprinted two edges on $n_{{\bm p},\sigma}$ play similar roles to Fermi surfaces. For this purpose, we consider the case when the Hamiltonian in the main system $H_{\rm sys}=H_0 +H_{\rm int}$ has the pairing interaction term,
\begin{equation}
H_{\rm int}= -U\sum_{\bm{p}, \bm{p}' , \bm{q}} a^\dagger_{\bm{p}+\bm{q}/2\up} a^\dagger_{-\bm{p}+\bm{q}/2\down}a_{-\bm{p}'+\bm{q}/2\down} a_{\bm{p}'+\bm{q}/2\up},
\label{eq.pairing}
\end{equation}
with $-U(<0)$ being an $s$-wave pairing interaction. As usual in cold atom physics \cite{Randeria1995}, we measure the interaction strength in terms of the $s$-wave scattering length $a_s$, which is related to the pairing interaction $-U$ as $4\pi a_s/m = -U/[1-U \sum_{\bm{p}}1/(2\ep_{\bm{p}})]$. In what follows, we focus on the weak-coupling BCS regime, to set $(a_s p_{\rm F})^{-1}=-1$, where $p_{\rm F}=\sqrt{2m\mu}$.
\par
The pairing interaction in Eq. (\ref{eq.pairing}) causes the superfluid instability, when the particle-particle scattering vertex $\chi(\bm{Q},\nu)$ develops a pole at $\nu=2\mu$ \cite{Kawamura2020, Kawamura2020JLTP}. Here, ${\bm Q}$ and $\nu$ are the center-of-mass momentum and the total energy of two particles participating in the Cooper channel, respectively. Within the random phase approximation in terms of $-U$, one has \cite{Kawamura2020, Kawamura2020JLTP}, in NESS, 
\begin{equation}
\chi(\bm{Q}, \nu=2\mu)=
{-U \over \displaystyle 1-{U \over 4\pi} \sum_{\eta, \zeta =\pm} 
{1 \over \xi^{\rm s}_{\bm{p},\bm{Q}}}{\rm Tan}^{-1}\left(\frac{\xi_{\bm{p},\bm{Q}}^{\eta,\zeta}}{2\gamma} \right)},
\label{eq.Thouless}
\end{equation}
where $\xi^{\rm s}_{\bm{p},\bm{Q}}=[\xi_{\bm{p}+\bm{Q}/2} + \xi_{-\bm{p}+\bm{Q}/2}]/2$, and $\xi_{{\bm p},{\bm Q}}^{\eta,\zeta}=\xi_{\bm{p} +\eta \bm{Q}/2} + \zeta \delta\mu$. 
\par
\begin{figure}[tb]
\centering
\includegraphics[width=80mm]{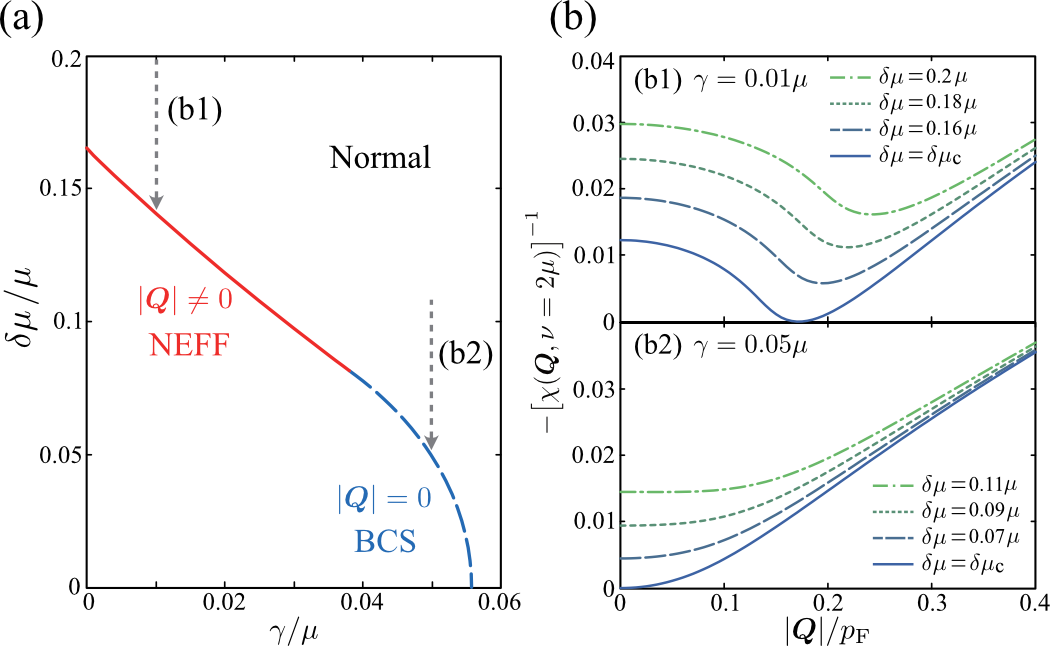}
\caption{(a) Calculated chemical potential difference $\delta\mu$ at the superfluid phase transition in the model driven-dissipative Fermi gas in Fig. \ref{fig1}(a). We take $T_{\rm env}=0$ and $(p_{\rm F}a_s)^{-1}=-1$. $\gamma$ is the damping rate. The main system exhibits the NEFF (BCS) superfluid instability on the solid (dashed) line. (b) Inverse particle-particle scattering vertex $[\chi(\bm{Q}, \nu=2\mu) ]^{-1}$, as a function of $\bm{Q}$. Upper (lower) panel shows the result along the path (b1) (path (b2)) in (a).}
\label{fig2} 
\end{figure}

Figure \ref{fig2}(a) shows the superfluid phase transition line in the $\gamma$-$\delta\mu$ plane, determined from the pole condition $\chi(\bm{Q},\nu=2\mu)^{-1}=0$. The region above this line is in the normal state, where the superfluid order is destroyed by strong incoherent pumping and decay of particles by the two reservoirs. When $\gamma/\mu\lesssim 0.04$, the pole condition is satisfied at ${\bm Q}\ne 0$ (see Fig. \ref{fig2}(b1)), indicating that the superfluid instability is associated with the Bose condensation of Cooper pairs with non-zero center of mass momentum. This situation is similar to the FF case in the thermal equilibrium state \cite{hu2006, Kinnunen2018}. Thus, the two edges in $n_{{\bm p},\sigma}$ is found to really work like two Fermi surfaces with different sizes $p_{\rm F1}$ and $p_{\rm F2}$. 
\par
Because the damping $\gamma$ makes the two-step structure in $n_{{\bm p},\sigma}$ obscure, the superfluid instability of this non-equilibrium FF-like state (NEFF) (${\bm Q}\ne 0$) changes to the BCS-type phase transition with ${\bm Q}=0$, when $\gamma/\mu\gesim 0.04$ (see Fig. \ref{fig2}(b2)). As one further increases $\gamma$, the two steps in $n_{{\bm p},\sigma}$ is completely smeared out and the overall structure becomes similar to the thermal equilibrium case at high temperatures. As a result, the main system is in the normal state, when $\gamma/\mu\gesim 0.056$ in Fig. \ref{fig2}(a).
\par
\par
\textit{Non-equilibrium superfluid phase}--\ We now enter the superfluid phase below the transition line in Fig. \ref{fig2}(a), by employing the following ansatz for the superfluid order parameter $\Delta(\bm{r},t)$:
\begin{equation}
\Delta(\bm{r},t)= \Delta_0 e^{-2i\mu t} e^{i \bm{Q}\cdot \bm{r}}. 
\label{eqOP}
\end{equation}
Here, $\Delta_0 \equiv U\sum_{\bm{p}}\braket{a_{-\bm{p}\down} a_{\bm{p}\up}}$ is taken to be positive real, without loss of generality. For simplicity, we do not consider the Larkin-Ovchinnikov type solution $\Delta(\bm{r},t)= \Delta_0 e^{-2i\mu t} \cos( \bm{Q}\cdot \bm{r})$ \cite{Larkin1964} in this letter. 
\par
In the thermal equilibrium state, $\Delta_0$ and ${\bm Q}$ of the FF superfluid order parameter can conveniently be determined from the minimization conditions for the free energy in terms of these quantities. However, this approach is not applicable to NESS, so that we take the following strategy: We derive the NESS gap equation from the self-consistent condition for $\Delta_0 = U\sum_{\bm{p}}\braket{a_{-\bm{p}\down} a_{\bm{p}\up}}$. Within the framework of the non-equilibrium Hartree-Fock-Bogoliubov approximation (NEHFB) \cite{Hanai2016, Hanai2017, Hanai2018}, this condition gives \cite{Kawamura2021}
\begin{equation}
1= \frac{U}{4\pi}\sum_{\bm{p}}\frac{1}{E_{\bm{p},\bm{Q}}}
\sum_{\eta, \zeta=\pm} {\rm Tan}^{-1}\left( \frac{E^{\eta,\zeta}_{\bm{p},\bm{Q}}}{2\gamma} \right),
\label{eq.gap}
\end{equation}
where $E^{\eta,\zeta}_{\bm{p},\bm{Q}}=E_{{\bm p},{\bm Q}}^\eta+\eta\xi_{{\bm p},{\bm Q}}^{\rm a}+\zeta\delta\mu$, with $\xi^{\rm a}_{\bm{p},\bm{Q}}=\big[ \xi_{\bm{p}+\bm{Q}/2} - \xi_{-\bm{p}+\bm{Q}/2}\big]/2$, and
\begin{equation}
E^{\pm}_{\bm{p},\bm{Q}}= \sqrt{(\xi^{\rm s}_{\bm{p},\bm{Q}})^2 +\Delta_0^2}\pm \xi^{\rm a}_{\bm{p},\bm{Q}}=E_{\bm{p},\bm{Q}} \pm \xi^{\rm a}_{\bm{p},\bm{Q}}.
\label{eq_Epq}
\end{equation}
The second equation to determine $(\Delta_0,{\bm Q})$ is obtained from the net current ${\bm J}_{\rm net}$ in the main system. In NEHFB \cite{Kawamura2021}, we have $\bm{J}_{\rm net} = \sum_{\sigma=\up, \down} \sum_{\bm{p}} \bigl[\bm{p}+\bm{Q}/2\bigr]n_{\bm{p}, \sigma}$, where
\begin{equation}
n_{{\bm p},\sigma}
=\frac{1}{2} -\frac{1}{4\pi} \sum_{\eta, \zeta=\pm}  {\rm Tan}^{-1}\left(\frac{E^{\eta,\zeta}_{\bm{p},\bm{Q}}}{2\gamma}\right)\left[ \eta+\frac{\xi^{\rm s}_{\bm{p},\bm{Q}}}{E_{\bm{p},\bm{Q}}} \right].
\label{eq.dist}
\end{equation}
According to the Bloch's theorem \cite{Bohm, OhashiMomoi}, ${\bm J}_{\rm net}$ must vanish in the thermal equilibrium state. However, this theorem does not work out of equilibrium. Indeed, one may consider the current-carrying superfluid state driven by external/boundary conditions \cite{Zagoskin2014,Samokhin2017}. In this paper, however, to simplify our discussions, we restrict our discussions to the case with ${\bm J}_{\rm net}=0$. That is, we solve the coupled gap equation (\ref{eq.gap}) with the vanishing-current condition, $\bm{J}_{\rm net}=0$, to self-consistently determine $(\Delta_0,\bm{Q})$.
\par
It can be shown that Eq. (\ref{eq.gap}) in the limit $\gamma\to 0$ has the same form as the FF gap equation in the thermal equilibrium state \cite{Fulde1964,Takada1969}. This means that the two edges at $p_{\rm F1}$ and $p_{\rm F2}$ in $n_{{\bm p},\sigma}$ work like large and small Fermi surfaces as in the FF case under an external magnetic field. On the other hand, when we further set ${\bm Q}=0$, Eq. (\ref{eq.gap}) has the same form as the ordinary BCS gap equation in the presence of an external magnetic field. From the knowledge about superconductivity, this coincidence means that the two-step structure in $n_{{\bm p},\sigma}$ does {\it not} promote the formation of Cooper-pairs (C) and (D) in Fig. \ref{fig1}(b), but rather suppresses the pair formation around $p_{\rm F}=\sqrt{2m\mu}$.
\par
\begin{figure}[tb]
\centering
\includegraphics[width=85mm]{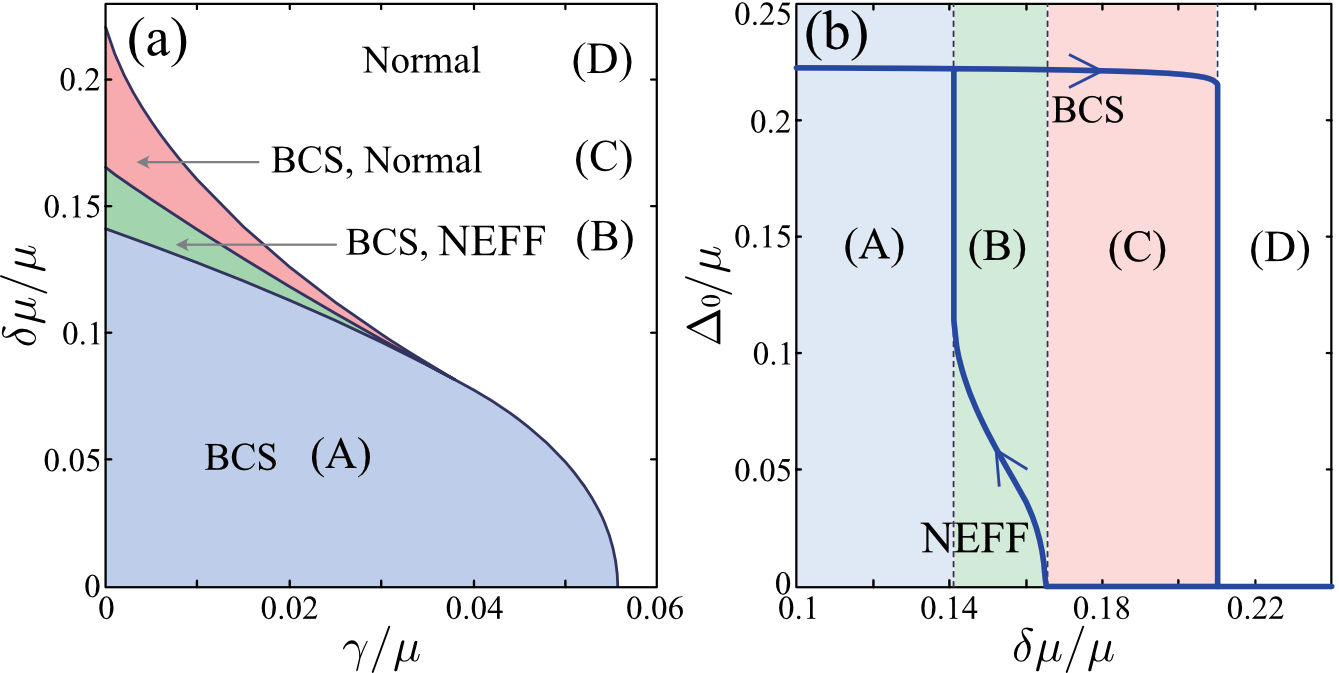}
\caption{(a) Phase diagram of a driven-dissipative Fermi gas. BCS: Non-equilibrium BCS-type superfluid ($\Delta_0>0$, $\bm{Q}=0$) state. NEFF: Non-equilibrium FF-like superfluid ($\Delta_0>0$, $\bm{Q}\neq 0$) state. (b) Hysteresis phenomenon in the regions (B) and (C) shown in panel (a). In panel (b), we take $\gamma \to +0$.}
\label{fig3} 
\end{figure}
\par
\textit{Phase diagram of driven-dissipative Fermi gas}--\ Figure \ref{fig3}(a) shows the steady-state phase diagram of a model driven-dissipative Fermi gas. We have confirmed that all the states appearing in this figure are (meta-)stable in the sense that the time evolution of a small {\it deviation} from each state always decays. (For more details, see Ref. \cite{Kawamura2021}.) As expected from Fig. \ref{fig2}(a), NEFF ($\Delta_0>0$ and ${\bm Q}\ne 0$) appears in the region (B), where the chemical potential difference $\delta\mu>0$ is large enough to produce a clear two-step structure in $n_{{\bm p},\sigma}$ but the damping $\gamma$ is not strong enough to smear out this structure. 
\par
We note that the BCS-type superfluid ($\Delta_0>0$ and ${\bm Q}=0$) is also stable in the region (B). This so-called bistability is a characteristic non-equilibrium phenomenon and has been observed in various systems \cite{Labouvie2015, Wang2018, Goldman1987}. This is quite different from the thermal equilibrium case, where the ground state is uniquely identified as the state with the lowest free energy. In the region (C), the bistability of the BCS state and the normal state occurs. 
\par
In the bistability regions (B) and (C), which state is realized would depend on how to reach these regions. When one varies $\delta\mu$ adiabatically, one expects the appearance of the hysteresis shown in Fig. \ref{fig3}(b): As $\delta\mu$ increases from $\delta\mu=0$, the BCS-type state ($\Delta_0>0$ and ${\bm Q}=0$) would be maintained both in the regions (B) and (C). As one decreases $\delta\mu$ from the region (D), on the other hand, the phase transition from the normal state to NEFF ($\Delta_0>0$ and ${\bm Q}\ne 0$) would occur at the boundary between (B) and (C).
\par
\begin{figure}[t]
\centering
\includegraphics[width=80mm]{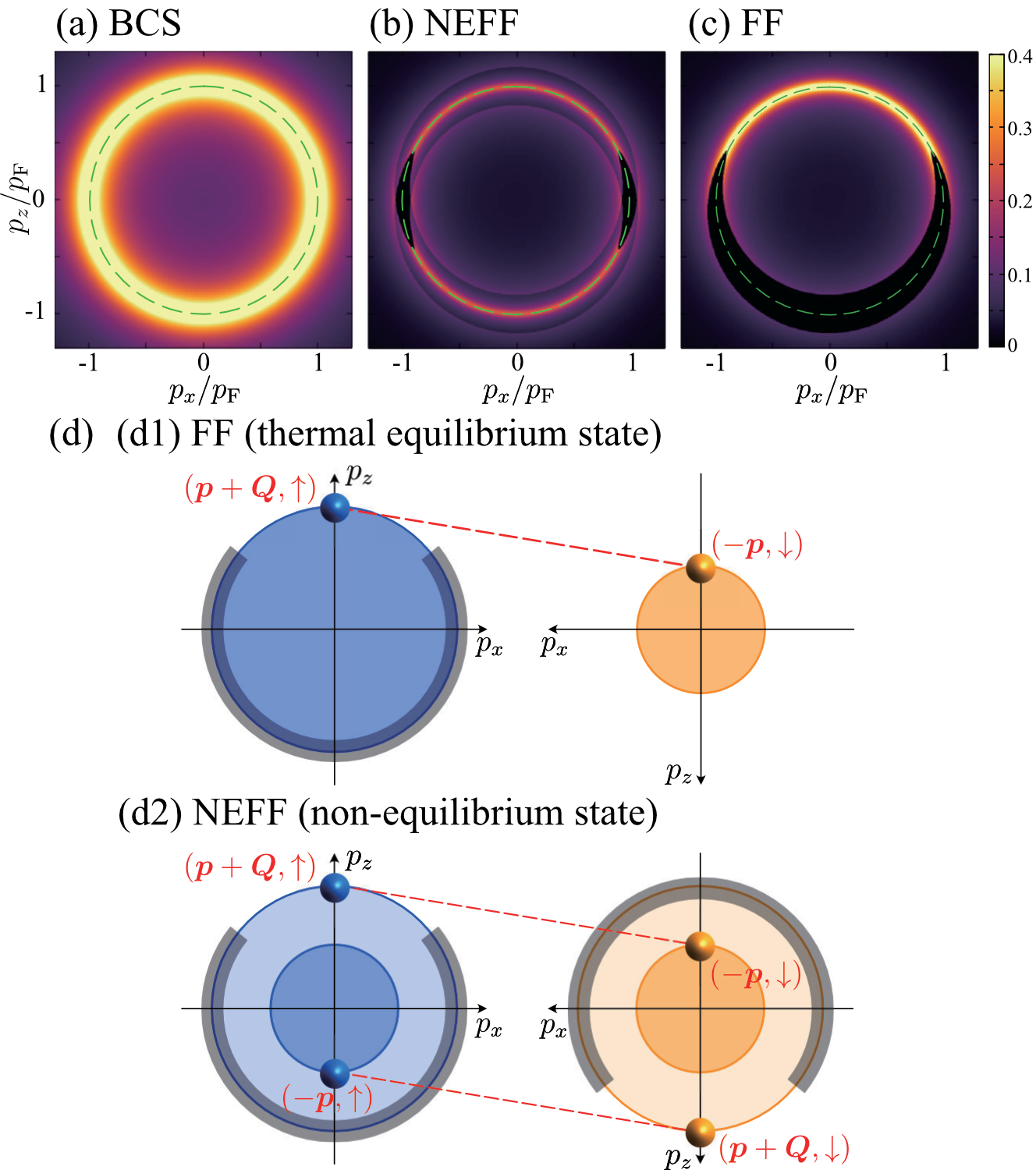}
\caption{Calculated pair amplitude $\braket{a_{-\bm{p}\down} a_{\bm{p}\up}}$. (a) BCS-type state. (b) NEFF state. We set $\delta\mu=0.15\mu$ and $\gamma\to +0$. For comparison, we also show in panel (c) the pair amplitude in the ordinary FF state realized in the thermal equilibrium state of a spin-polarized Fermi gas. The dashed lines in panels (a)-(c) show the positions at $p=\sqrt{2m\mu}$. In panels (b) and (c), the direction of $\bm{Q}$ is chosen in the $p_z$ direction. (d) Schematic pictures of pair-formation: (d1) FF state. (d2) NEFF state. Fermions in the shaded regions (blocking regions) do not contribute to the pair formation.}
\label{fig4} 
\end{figure}

To see the difference between the BCS-type and NEFF states in Fig. \ref{fig3}(a), we compare in Figs. \ref{fig4}(a) and (b) their pair amplitudes, both of which are commonly given by
\begin{equation}
\braket{a_{-\bm{p}\down}a_{\bm{p}\up}} = -\frac{1}{4\pi} \sum_{\eta, \zeta=\pm} {\rm Tan}^{-1}\left(\frac{E^{\eta,\zeta}_{\bm{p},\bm{Q}}}{2\gamma}\right) \frac{\Delta_0}{E_{\bm{p},\bm{Q}}}.
\label{pair_amp}
\end{equation}
While the pair amplitude is isotropic in the BCS-type state (see Fig. \ref{fig4}(a)), Fig. \ref{fig4}(b) shows that it vanishes around the ``equator" of the Fermi sphere in NEFF, when ${\bm Q}$ points to the $p_z$ direction. 
\par
We emphasize that this structure of the NEFF pair amplitude is also different from the FF case in the thermal equilibrium state shown in Fig. \ref{fig4}(c): The vanishing region (which is also referred to the blocking region in the superconductivity literature) spreads over the lower hemisphere. Noting that NEFF may be viewed as a mixture of two FF states with ${\bm Q}$ and $-{\bm Q}$ as shown in Figs. \ref{fig4}(d), we find that their blocking regions give the vanishing pair amplitude around the equator in Fig. \ref{fig4}(b). 
We briefly note that the difference of pair amplitude $\langle a_{-{\bm p}\downarrow}a_{{\bm p}\uparrow}\rangle$ between NEFF and FF does {\it not} reflect the order parameter $\Delta_0=U\sum_{\bm{p}} \langle a_{-{\bm p}\downarrow}a_{{\bm p}\uparrow}\rangle$ in the weak damping limit $\gamma\to+0$. In this limit, thus, they follow the same gap equation.
\par
So far, we have only discussed stable solutions of the non-equilibrium gap equation (\ref{eq.gap}) under the vanishing current condition $\bm J_{\rm net}=0$. Here, we briefly comment on {\it unstable} solutions: One is a gapless uniform superfluid state (${\bm Q}=0$), being accompanied by the Cooper pairs (C) and (D) in Fig. \ref{fig1}(b). The other is another FF-like state where $\bm{Q}$ is nonzero but smaller than the stable NEFF state. We note that similar unstable solutions are known as the Sarma(-Liu-Wilczek) state \cite{Sarma1963, Liu2003}, as well as the saddle point \cite{hu2006} state in spin- and mass-imbalanced Fermi gases, respectively.
\par
\par
\par
\textit{Summary and future work}--\ We have proposed an idea to process the Fermi momentum distribution $n_{{\bm p},\sigma}$, by using reservoirs with different chemical potentials. We considered a driven-dissipative two-component Fermi gas, to show that the pumping and decay of fermions by two reservoirs can imprint a two-edge structure on $n_{{\bm p},\sigma}$. We showed that, in the presence of an attractive pairing interaction between fermions, the two edges work like two Fermi surfaces with different Fermi momenta $p_{\rm F1}$ and $p_{\rm F2}$, leading to a non-uniform FF-like superfluid state in the non-equilibrium steady state. We clarified the similarity and difference between this unconventional pairing state and the FF state known in superconductivity under an external magnetic field. 
\par
Our proposed FS-enginearing can be applied to a variety of situations. In (optical) lattice systems, the combination of the band structure and our technique may trigger unconventional ordered phases, such as non-equilibrium spin- and charge-density wave-like states. Since many-fermion systems are sensitive to their FSs, the proposed FS-engineearing would contribute to further exploration for unknown many-body quantum phenomena associated with FSs. 
\par
\par
We thank D. Kagamihara and K. Furutani for discussions. T.K. was supported by MEXT and JSPS KAKENHI Grant-in-Aid for JSPS fellows Grant No.JP21J22452. R.H. was upported by an appointment to the JRG Program at the APCTP through the Science and Technology Promotion Fund and Lottery Fund of the Korean Governmen. Y.O. was supported by a Grant-in-aid for Scientific Research from MEXT and JSPS in Japan (No.JP18K11345, No.JP18H05406, and No.JP19K03689).
\par
\par


\begin{thebibliography}{99}
\bibitem{Kondobook} For a review, see, {\it Fermi Surface Effects} edited by J. Kondo, and A. Yoshimori (Springer, Berlin, 1988). 
\bibitem{Anderson} P. W. Anderson, Phys. Rev. Lett. {\bf 18}, 1049 (1967).
\bibitem{Cooper1956} L. N. Cooper, Phys. Rev. {\bf 104}, 1189 (1956).
\bibitem{Grunerbook} G. Gr\"uner, {\it Density Waves in Solids}, (CRC Press, NY, 1994).
\bibitem{Fulde1964} P. Fulde and R. A. Ferrell, Phys. Rev. {\bf 135}, A550 (1964).
\bibitem{Takada1969} S. Takada and T. Izuyama, Prog. Theor. Phys. {\bf 41}, 635 (1969).
\bibitem{Chu1970} C. W. Chu, T. F. Smith, and W. E. Gardner, Phys. Rev. B {\bf 1}, 214 (1970). 
\bibitem{Prieto2006} A. Rodriguez-Prieto, A. Bergara, V. M. Silkin, and P. M. Echenique, Phys. Rev. B, {\bf 74}, 172104 (2006).
\bibitem{Glazyrin2013} K. Glazyrin, L. V. Pourovskii, L. Dubrovinsky, O. Narygina, C. McCammon, B. Hewener, V. Sch\"{u}nemann, J. Wolny, K. Muffler, A. I. Chumakov, W. Crichton, M. Hanfland, V. B. Prakapenka, F. Tasnadi, M. Ekholm, M. Aichhorn, V. Vildosola, A. V. Ruban, M. I. Katsnelson, and I. A. Abrikosov, Phys. Rev. Lett. {\bf 110}, 117206 (2013).
\bibitem{Xiang2015} Z. J. Xiang, G. J. Ye, C. Shang, B. Lei, N. Z. Wang, K. S. Yang, D. Y. Liu, F. B. Meng, X. G. Luo, L. J. Zou, Z. Sun, Y. Zhang, X. H. Chen, Phys. Rev. Lett. {\bf 115}, 186403 (2015).
\bibitem{Kang2015} D. Kang, Y. Zhou, W. Yi, C. Yang, J. Guo, Y. Shi, S. Zhang, Z. Wang, C. Zhang, S. Jiang, A. Li, K. Yang, Q. Wu, G. Zhang, L. Sun, Z. Zhao, Nature Communications {\bf 6}, 7804 (2015).
\bibitem{Gonnelli2016} R. S. Gonnelli, D. Daghero, M. Tortello, G. A. Ummarino, Z. Bukowski, J. Karpinski, P. G. Reuvekamp, R. K. Kremer, G.
Profeta, K. Suzuki, and K. Kuroki, Sci. Rep. {\bf 6}, 26394 (2016).
\bibitem{Testardi1970} L. R. Testardi and J. H. Condon, Phys. Rev. {\bf B} 1, 3928 (1970).
\bibitem{Martins1978} J. M. V. Martins, F. P. Missell, and J. R. Pereira, Phys. Rev. B {\bf 17}, 4633 (1978).
\bibitem{Burganov2016} B. Burganov, C. Adamo, A. Mulder, M. Uchida, P. D. C. King, J. W. Harter, D. E. Shai, A. S. Gibbs, A. P. Mackenzie, R. Uecker, M. Bruetzam, M. R. Beasley, C. J. Fennie, D. G. Schlom, K. M. Shen, Phys. Rev. Lett. {\bf 116}, 197003 (2016).
\bibitem{Hsu2016} Y. T. Hsu, W. Cho, A. F. Rebola, B. Burganov, C. Adamo, K. M. Shen, D. G. Schlom, C. J. Fennie, and E. A. Kim, Phys. Rev. B {\bf 94}, 045118 (2016).
\bibitem{Sunko2019} V. Sunko, E. A. Morales, I. Markovi\'{c}, M. E. Barber, D. Milosavljevi\'{c}, F. Mazzola, D. A. Sokolov, N. Kikugawa, C. Cacho, P. Dudin, H. Rosner, C. W. Hicks, P. D. C. King, and A. P. Mackenzie, npj Quantum Materials {\bf 4}, 46 (2019).
\bibitem{Armitage2002} N. P. Armitage, F. Ronning, D. H. Lu, C. Kim, A. Damascelli, K. M. Shen, D. L. Feng, H. Eisaki, Z.-X. Shen, P. K. Mang, N. Kaneko, M. Greven, Y. Onose, Y. Taguchi, and Y. Tokura, Phys. Rev. Lett. {\bf 88}, 257001 (2002).
\bibitem{Kusunose2003} H. Kusunose and T. M. Rice, Phys. Rev. Lett. {\bf 91}, 186407 (2003).
\bibitem{Kaminski2006} A. Kaminski, S. Rosenkranz, H. M. Fretwell, M. R. Norman, M. Randeria, J. C. Campuzano, J-M. Park, Z. Z. Li, and H. Raffy, Phys. Rev. B {\bf 73}, 174511 (2006).
\bibitem{Liu2010} C. Liu, T. Kondo, R. M. Fernandes, A. D. Palczewski, E. D. Mun, N. Ni, A. N. Thaler, A. Bostwick, E. Rotenberg, J. Schmalian, S. L. Bud’ko, P. C. Canfield, A. Kaminski, Nature Physics {\bf 6}, 419 (2010).
\bibitem{Shi2017} X. Shi, Z.-Q. Han, X.-L. Peng, P. Richard, T. Qian, X.-X. Wu, M.-W. Qiu, S. C. Wang, J. P. Hu, Y.-J. Sun, H. Ding, Nature Communications {\bf 8}, 14988 (2017).
\bibitem{Beaulieu2020} S. Beaulieu, S. Dong, N. Tancogne-Dejean, M. Dendzik, T. Pincelli1, J. Maklar, R. P. Xian, M. A. Sentef, Martin Wolf, Angel Rubio, L. Rettig, and Ralph Ernstorfer, arXiv:2003.04059.
\bibitem{Kirby2020} R. J. Kirby, L. Muechler, S. Klemenz, C. Weinberg, A. Ferrenti, M. Oudah, D. Fausti, G. D. Scholes, and L. M. Schoop, Phys. Rev. B {\bf 103}, 205138 (2021).
\bibitem{Pothier1997} H. Pothier, S. G\'{u}eron, N. O. Birge, D. Esteve, and M. H. Devoret, Phys. Rev. Lett. {\bf 79}, 3490 (1997).
\bibitem{Anthore2003} A. Anthore, F. Pierre, H. Pothier, and D. Esteve, Phys. Rev. Lett. {\bf 90}, 076806 (2003).
\bibitem{Chen2009} Y.-F. Chen, T. Dirks, G. Al-Zoubi, N. O. Birge, and N. Mason, Phys. Rev. Lett. {\bf 102}, 036804 (2009).
\bibitem{Brantut2012} J.-P. Brantut, J. Meineke, D. Stadler, S. Krinner, and T. Esslinger, Science {\bf 337}, 1069 (2012).
\bibitem{Krinner2015} S. Krinner, D. Stadler, D. Husmann, J.-P. Brantut, and T. Esslinger, Nature (London) {\bf 517}, 64 (2015).
\bibitem{Husmann2015} D. Husmann, S. Uchino, S. Krinner, M. Lebrat, T. Giamarchi, T. Esslinger, and J.-P. Brantut, Science {\bf 350}, 1498 (2015).
\bibitem{Krinner2017} S. Krinner, T. Esslinger, and J.-P. Brantut, J. Phys.: Condens. Matter {\bf 29}, 343003 (2017).
\bibitem{Matsuda2007} Y. Matsuda and H. Shimahara, J. Phys. Soc. Jpn. {\bf 76}, 051005 (2007).
\bibitem{Kinnunen2018} J. J. Kinnunen, J. E. Baarsma, J.-P. Martikainen, and P. T\"{o}rm\"{a}, Rep. Prog. Phys. {\bf 81}, 046401 (2018).
\bibitem{hu2006} H. Hu and X.-J. Liu, Phys. Rev. A {\bf 73}, 051603(R) (2006).
\bibitem{Liao2010} Y. Liao, A. S. C. Rittner, T. Paprotta, W. Li, G. B. Partridge, R. G. Hulet, S. K. Baur, and E. J. Mueller, Nature (London) {\bf 467}, 567 (2010).
\bibitem{Chevy2010} F. Chevy and C. Mora, Rep. Prog. Phys. {\bf 73}, 112401 (2010).
\bibitem{Casalbuoni2004} R. Casalbuoni and G. Nardulli, Rev. Mod. Phys. {\bf 76}, 263 (2004).
\bibitem{Doh2006} H. Doh, M. Song, and H.-Y. Kee, Phys. Rev. Lett. {\bf 97}, 257001 (2006).
\bibitem{Vorontsov2009} A. B. Vorontsov, Phys. Rev. Lett. {\bf 102}, 177001 (2009).
\bibitem{He2018} L. He, H. Hu, and X.-J. Liu, Phys. Rev. Lett. {\bf 120}, 045302 (2018).
\bibitem{Zheng2015} Z. Zheng, C. Qu, X. Zou, and C. Zhang, Phys. Rev. A {\bf 91}, 063626 (2015).
\bibitem{Zheng2016} Z. Zheng, C. Qu, X. Zou, and C. Zhang, Phys. Rev. Lett. {\bf 116}, 120403 (2016).
\bibitem{Nocera2017} A. Nocera, A. Polkovnikov, and A. E. Feiguin, Phys. Rev. A {\bf 95}, 023601 (2017).
\bibitem{Kawamura2020JLTP} T. Kawamura, D. Kagamihara, R. Hanai, and Y. Ohashi, J Low Temp Phys {\bf 201}, 41-48 (2020).
\bibitem{Kawamura2020} T. Kawamura, R. Hanai, D. Kagamihara, D. Inotani, Y. Ohashi, Phys. Rev. A {\bf 101}, 013602 (2020). 
\bibitem{Hanai2016} R. Hanai, P. B. Littlewood, and Y. Ohashi, J. Low Temp. Phys. {\bf 183}, 127 (2016).
\bibitem{Hanai2017} R. Hanai, P. B. Littlewood, and Y. Ohashi, Phys. Rev. B {\bf 96}, 125206 (2017).
\bibitem{Hanai2018} R. Hanai, P. B. Littlewood, and Y. Ohashi, Phys. Rev. B {\bf 97}, 245302 (2018).
\bibitem{Stefanucci2013} G. Stefanucci and R. van Leeuwen, {\it Nonequilibrium Many-Body Theory of Quantum Systems: A Modern Introduction} (Cambridge University Press, Cambridge, UK, 2013).
\bibitem{Randeria1995} M. Randeria, in {\it Bose-Einstein Condensation}, edited by A. Griffin, D. W. Snoke, and S. Stringari (Cambridge University Press, UK, 1995), pp. 355-392.
\bibitem{Larkin1964} A. I. Larkin and Y. N. Ovchinnikov, Zh. Eksp. Teor. Fiz. {\bf 47}, 1136 (1964) [Sov. Phys. JETP {\bf 20}, 762 (1965)].
\bibitem{Kawamura2021} T. Kawamura, R.Hanai, Y. Ohashi, companion paper, in preparation.
\bibitem{Bohm} D. Bohm, Phys. Rev. {\bf 75}, 502 (1949).
\bibitem{OhashiMomoi} Y. Ohashi and T. Momoi, J. Phys. Soc. Jpn. {\bf 65}, 3254 (1996).
\bibitem{Zagoskin2014} A. Zagoskin, {\it Quantum Theory of Many-Body Systems} (Springer, New York, 2014).
\bibitem{Samokhin2017} K. V. Samokhin and B. P. Truong, Phys. Rev. B {\bf 96}, 214501 (2017).
\bibitem{Goldman1987} V. J. Goldman, D. C. Tsui, and J. E. Cunningham, Phys. Rev. Lett. {\bf 58}, 1256 (1987).
\bibitem{Labouvie2015} R. Labouvie, B. Santra, S. Heun, S. Wimberger, and H. Ott, Phys. Rev. Lett. {\bf 115}, 050601 (2015).
\bibitem{Wang2018} Y.-P. Wang, G.-Q. Zhang, D. Zhang, T.-F. Li, C.-M. Hu, and J.-Q. You, Phys. Rev. Lett. {\bf 120}, 057202 (2018).
\bibitem{Sarma1963} G. Sarma, J. Phys. Chem. Solids {\bf 24}, 1029 (1963).
\bibitem{Liu2003} W. V. Liu and F. Wilczek, Phys. Rev. Lett. {\bf 90}, 047002 (2003).
\end{thebibliography}
\end{document}